# Control the high-order harmonics cutoff through the combination of chirped laser and static electric field

Yang Xiang[1, 2, 3] Yueping Niu [1] Shangqing Gong [1]
1 State Key Laboratory of High Field Laser Physics, Shanghai Institute of Optics and Fine Mechanics, Chinese Academy of Science, Shanghai 201800, China
2 Graduate School of Chinese Academy of Sciences, Beijing 100049 China
3 College of Computer Science and Technology, Henan Polytechnic University, Jiaozuo 454000 China

**Abstract:** The high harmonic generation from atoms in the combination of chirped laser pulse and static field is theoretically investigated. For the first time, we explore a further physical mechanism of the significant extension of high harmonic generation cutoff based on three-step model. It is shown that the cutoff is substantially extended due to the asymmetry of the combined field. If appropriate parameters are chosen, the cutoff of high harmonic generation can reach $I_p + 42U_p$. Furthermore, an ultrabroad super-continuum spectrum can be generated. When the phases are properly compensated for, an isolated 9 attosecond pulse can be obtained.

It is well known that high harmonics are generated when atoms or molecules are irradiated by intense laser field. In recent years, high harmonic generation (HHG) has become a very interesting topic of laser-atom interaction [1-3] because of a very important application of attosecond (as) pulses generation. Usually, there are two techniques to generate attosecond pulses [4]. One is to use few-cycle laser pulse, then the cutoff is continuous and by spectrally selecting it a single attosecond pulse can be obtained [5]. The other is to use multi-cycle intense infrared pulse, then select many harmonics in the plateau and an attosecond pulse train can be generated [6]. Many control ways have been studied in order to gain an isolated attosecond pulse [7-9]. In the presence of an intense static electric field, one can get a pulse of 220 attosecond [10]. Using a chirped few-cycle laser field, a pulse of 108 attosecond can be generated [11]. From the works mentioned above, one may want to know what would happen if the combination of chirped laser field and a static electric field is adopted in the HHG process. Based on this idea, we now investigate the HHG spectrum of atoms in this combined field.

We use one-dimensional time-dependent Schrödinger equation (TDSE) to describe the interaction between an atom and the combination of chirped laser field and the static electric field in the single active electron and dipole approximations (The atom units are used in all equations in this paper, unless otherwise mentioned.):

$$i\frac{\partial\psi(x,t)}{\partial t} = \left[-\frac{1}{2}\frac{\partial^2}{\partial x^2} + V(x) - xE(t)\right]\psi(x,t), \tag{1}$$

where the Coulomb potential $V(x)$ is represented by the 'soft-core' potential which can be expressed as: $V(x) = -1/\sqrt{\rho^2 + x^2}$ .Here, $\rho$ is the softening parameter. We choose $\rho$ =0.6957 which corresponds to the ionization potential ($I_p$) of 24.6 eV for the ground state of the helium atom. The combined field is described as

$$E(t) = F\left[ f(t)\cos(\omega t + \delta(t)) - \alpha \right], \tag{2}$$

where $F$, $f(t)$, $\omega$ and $\delta(t)$ is the amplitude, envelope, angular frequency and time profile of the carrier envelope phase (CEP) of the laser field, and $\alpha$ is the ratio between the amplitude of the static and laser field. If $\alpha = 0$ and $\delta(t) = 0$, $E(t)$ is of symmetry. However the symmetry will be destroyed if $\alpha \neq 0$ or $\delta(t) \neq 0$. In all of our calculations, F is $5.0 \times 10^{14}$ W/cm$^2$, $\omega$ is 0.057 a.u. (corresponds to wavelength $\lambda$ =800 nm) and $\alpha$ is equal to 0.4 which is approximate to that in Ref. [10]. Although such a high static field can hardly be experimentally achieved nowadays, a low-frequency laser field (such as $CO_2$ lasers) can be used instead [12]. The envelope of the laser field $f(t)$ has a type of Gaussian with 5 femtosecond full width at half maximum. The CEP considered in this paper is $\delta(t) = -\beta \tanh((t - t_0)/\tau)$ [11]. The parameters $\beta$, $t_0$ and $\tau$ are used to control the chirp form. Due to the recent advancement of comb laser technology, it is highly likely that such a time-varying CEP can be achieved in near future [11, 13, 14]. In our work, $\tau$ is chosen to be 200 a.u..

In order to find how the combined field impacts on the HHG, we consider the HHG of helium atom in four field cases: (a) chirp-free laser field; (b) combination of chirp-free laser field and static electric field; (c) chirped laser field; (d) combination of chirped laser field and static electric field. The parameters $\beta$ and $t_0$ for the chirped laser field in (c) and (d) are 6.25 and $\tau$/7.0. These fields are shown in Fig.1.

The HHG spectrums of the four cases are shown in Fig.2. The cutoff of the spectrums for (b), (c) and (d) is about 171st, 241st and 551st-order harmonic respectively, and each of them is much higher than the well-known value of $I_p + 3.17U_p$ (about 77th order harmonic, just as shown in Fig.2(a)), where $U_p (= F^2/(4\omega^2))$is the ponderomotive energy. To our surprise, the cutoff of HHG in (d) is much higher than that in (b) or (c). Furthermore, an ultra-broad super-continuum spectrum which covers about 450 orders harmonics appears in the case of combination of chirped laser field and static electric field.

Generally, the physical origin of the HHG from linearly polarized laser field can be qualitatively understood in the framework of semi-classical model which is called three-step-method (STM) [15]: first, the electron tunnels through the barrier formed by the Coulomb potential and the laser field; next, it oscillates almost freely in the laser field; finally, it may return back and recombine with the parent ion. During the recombination, a photon is emitted.

The returning kinetic energies of the electron in different laser fields are shown in Fig. 2. For comparison, the ionization potential is added. It is apparent that the trajectories of the electron in these field cases are quite different. Usually, there are two dominant quantum paths with different emission times contributing to each harmonic in each half optical cycle (O.C.) [10]: the positive slope section (short path) and negative slope section (long path) in returning kinetic energy map. In Fig.2 (b), due to the effect of static electric field, only the short paths contribute to the harmonics above $I_p + 0.62U_p$, and the maximum returning kinetic energy (MRKE) is up to $8.0U_p$ (about 154 harmonics). That is to say, the static electric field can eliminate some long paths of the electron. In Fig.2 (c), the long paths of the electron are not eliminated as apparently as those in Fig.2(b), but the MRKE is up to $11.7U_p$ (about 225 harmonics) because of the chirped laser field. However, when the combination of the chirped laser and static electric field is applied, the electron returns within less than 1.5 optical cycles, and for harmonics above $I_p + 2.1U_p$, only the short path contributes to the harmonic generation, just as Fig.2 (d) displays. The MRKE is up to $27.5U_p$ (about 530 harmonics) which is much higher than the summation of that in Fig.2 (b) and Fig.2 (c). If we mark the maximum peak (cutoff-energy) and the peak just below it with M and N respectively, we can see clearly that it is the difference between the peak M and N that contributes to the continuum spectrum, as is mentioned in Ref. [8]. Just as reviewed in the introduction, the continuous spectrum is beneficial to generate single attosecond pulse. This will be discussed later.

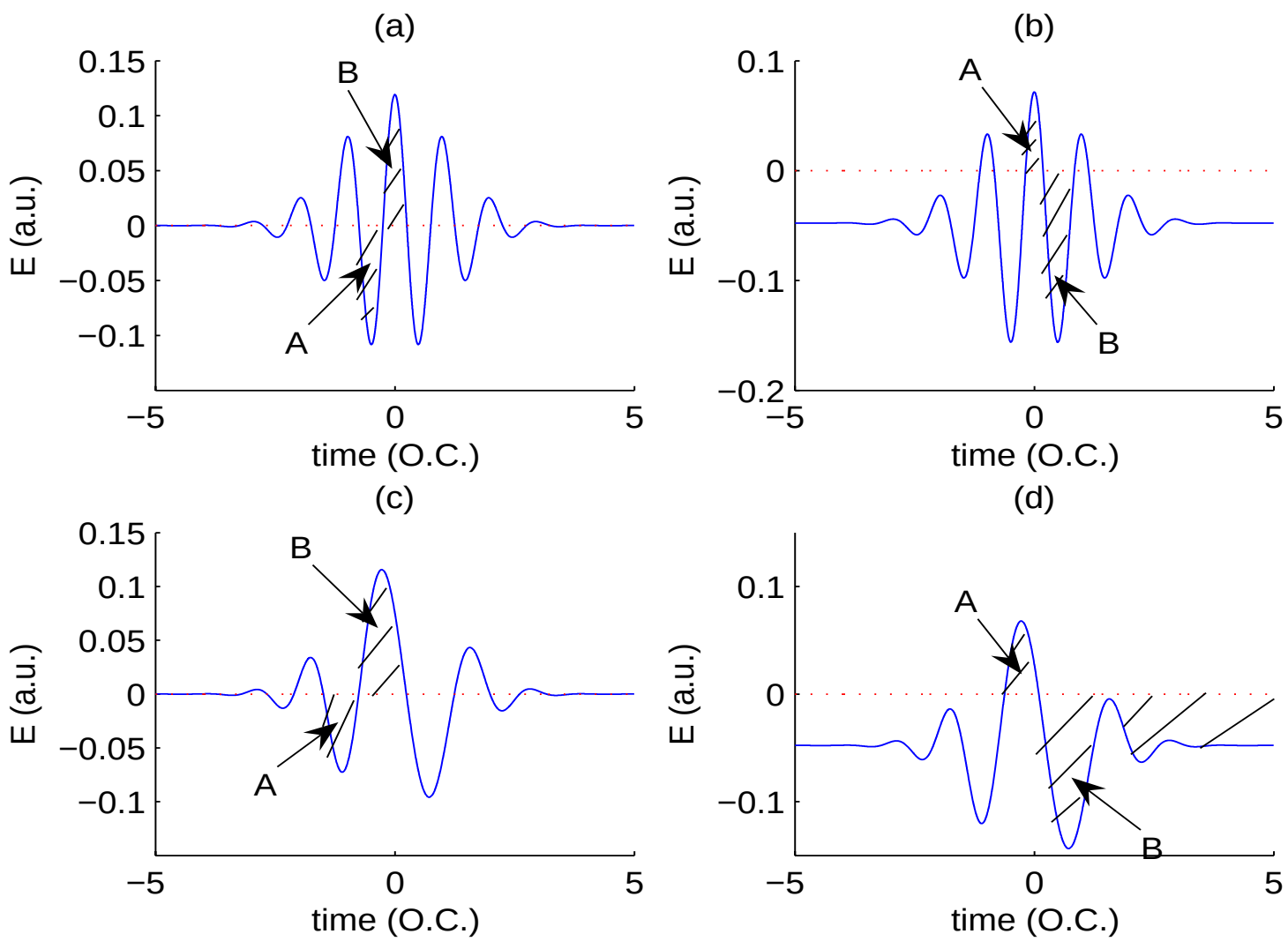

**Figure** 1 (a) chirp-free laser field; (b) combination of chirp-free laser field and static electric field; (c) chirped laser field; (d) combination of chirped laser field and static electric field.

In order to explore the extremely extended cutoff, we look into two processes in TSM: (i) once the electron tunnels, it accelerates in the electric field; (ii) when the electric field inverses, the electron then slows down and turn back to the parent ion. If the impulse in process (i) is weak, the kinetic energy the electron gains is small, as well as the returning kinetic energy. However, if the impulse in process (i) is strong, the impulse in process (ii) should be much stronger than the impulse in process (i), otherwise, the electron can not return to the parent ion. So, in order to gain a large returning kinetic energy, both the impulse in process (i) and the difference between the two impulses should be large. Based on this, one can easily find the two impulses which contribute to the MRKE. These impulses are represented by the shadow areas which marked with A and B respectively in each field of Fig.1. Compared with the returning energy map in Fig. 2, we find that the electron of the maximum energy just returns in area B. From Fig.1, we can see that there is not so much difference of impulse A in the four fields. Nevertheless, due to the static electric field or the chirp, the symmetry of the field is broken. Thus, the difference of impulse B and impulse A enlarges, so the cutoff of case (b) or (c) is extended, compared with that of case (a). For the combination of the chirped laser and the static field, the symmetry is broken further, and hence there is a greater difference between impulse B and impulse A. As a result, the cutoff is extended significantly. From the analysis above, we conclude that proper asymmetry of the field can lead to the extension of the HHG cutoff extremely.

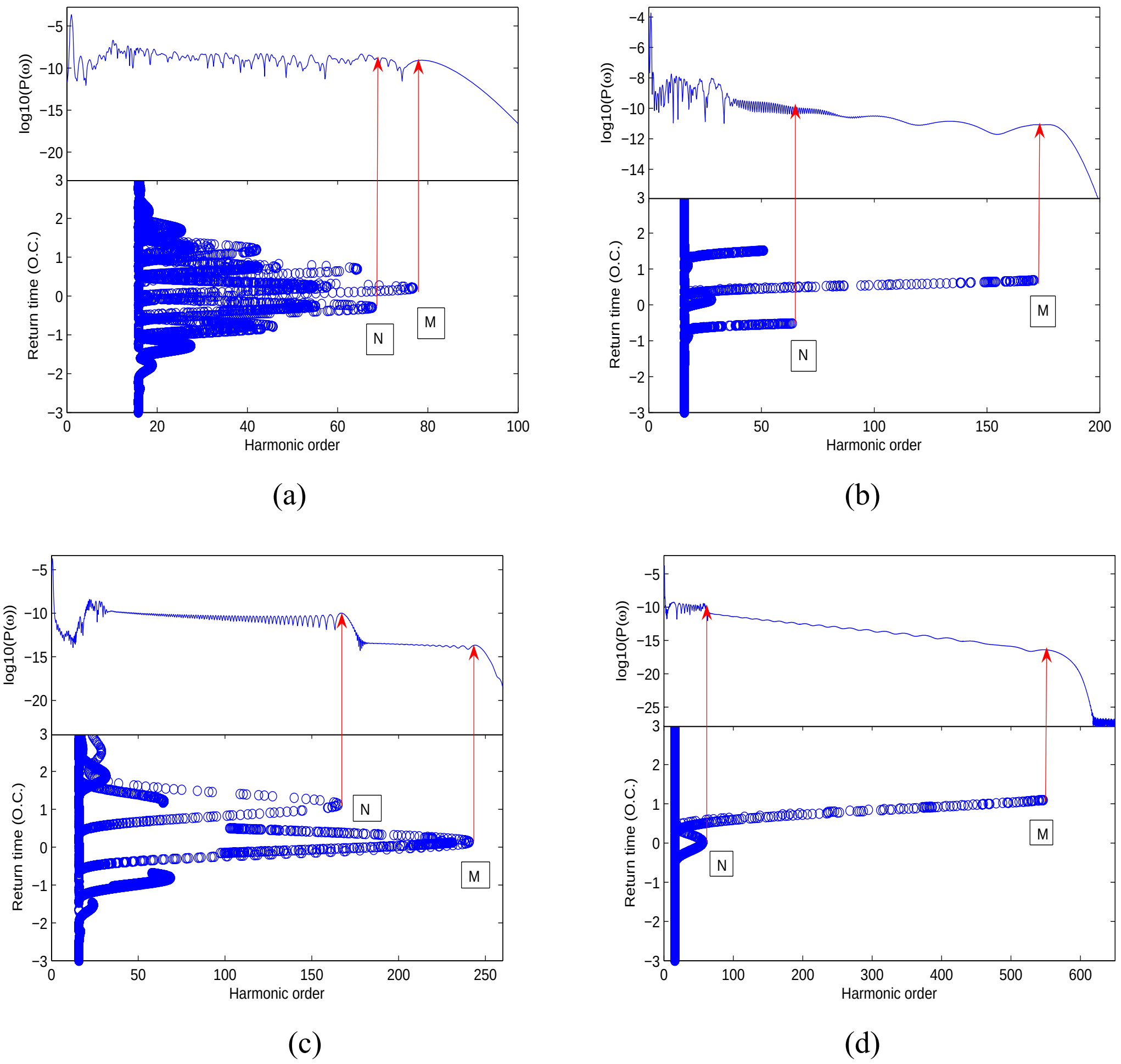

**Figure 2** Comparison of the returning kinetic energy of the electron and HHG spectrums in different field conditions: (a) chirp-free laser field; (b) combination of chirp–free laser and static electric field; (c) chirped laser field; d) combination of chirped laser and static electric field. The parameters are same to those in Fig. 1.

Since the cutoff of HHG can be extremely extended in the presence of chirped laser and static electric field, it is instructive to find out how the chirp parameters impact on the HHG cutoff. The variation of the MRKE of electron with the chirp parameters $\beta$ and $t_0$ is obtained by TSM, and the result is shown in Fig.3. Apparently the optimal MRKE of electron is up to $42U_p$ at $\beta = 9.0$ and $t_0 = \tau / 8.0$. The HHG spectrum for these parameters is presented in Fig.4. One can find that the continuum spectrum is broadened to 700 order harmonics.

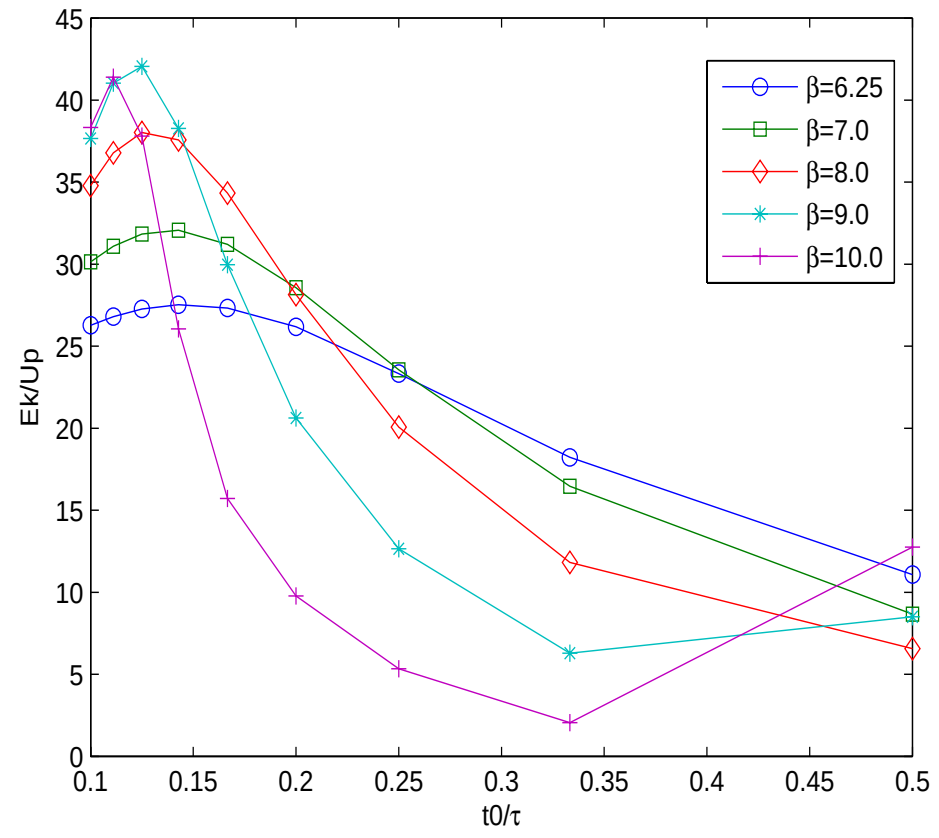

**Figure 3** The maximum returning energy of electron as a function of $\beta$ and $t_0$.

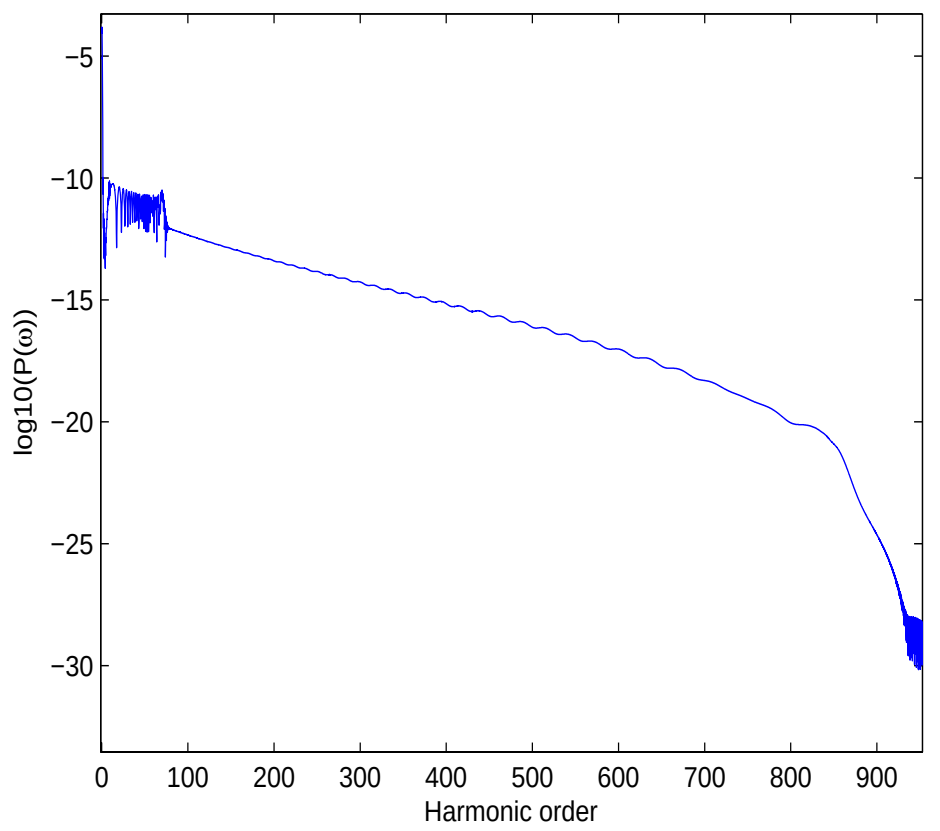

**Figure 4** HHG spectrum for the combination of chirped laser and static field. The chirp parameters are: $\beta = 9.0$, $t_0 = \tau / 8.0$.

Now, we consider the attosecond pulse generation from HHG of Fig.4. By adding a frequency window with a bandwidth of 50 order harmonics to the super-continuum spectrum and making an inverse Fourier transformation, an isolated attosecond pulse is generated without any phase compensation. If we move this frequency window along harmonic order axis, the duration of the attosecond pulse changes a little, though the intensity decreases. That is to say, the pulse duration is not sensitive to the position of the frequency window. The temporal profiles of the attosecond pulses are shown in Fig.5 (a). The shortest pulse duration we show in this figure is 56 attosecond. One may think that if we enlarge the width of the frequency window, the attosecond pulse duration will shorten further. Actually, due to the phase mismatch, the pulse duration lengthens but rather shortens, as shown in Fig.5 (b) (compared with the red dot line in (a)). However, if the phase mismatch over such an ultra-broad continuum spectral can be properly compensated for, an isolated 9 attosecond pulse with a clean temporal profile (Fig.5(c)) could be theoretically obtained.

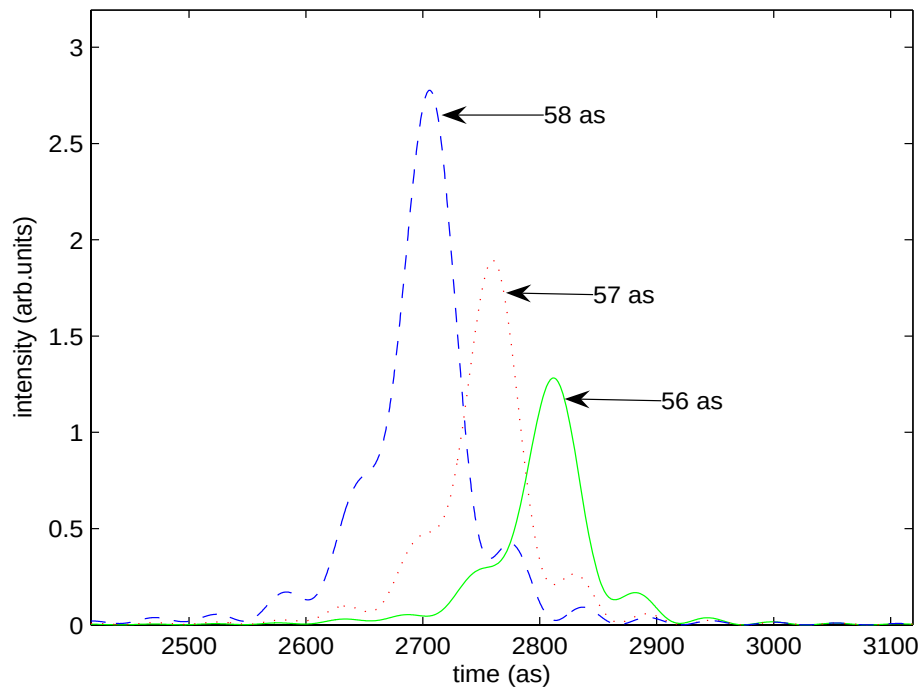

(a)

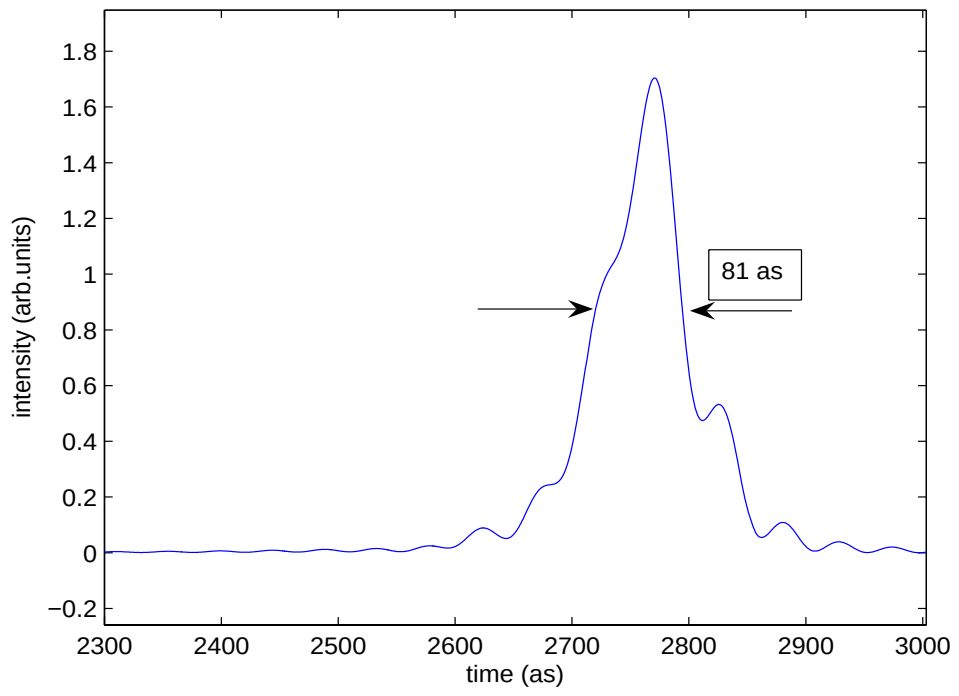

(b)

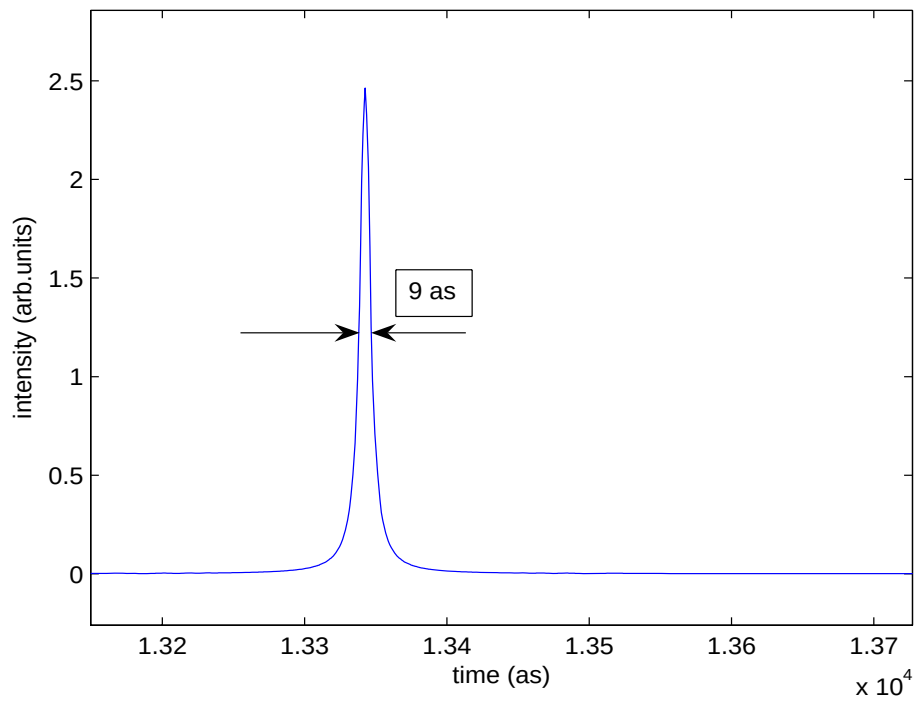

(c)

**Figure 5** The temporal profiles of the attosecond pulses generated from the continuum spectrum without (a), (b) and with (c) phase compensation. The harmonics order used in (a) are: 220th- 270th (blue dash line), 240th- 290th (red dot line), 260th- 310th (green solid line).The harmonics order used in (b) is: 240th-300th.The harmonics order used in (c) is: 220th-700th.

In conclusion, the HHG of helium atom in the combination of chirped laser field and static electric field has been investigated. We demonstrated that proper asymmetry of the field lead to the significant extension of the HHG cutoff. If appropriate parameters are used, an ultra-broad continuum spectrum which covers about 700 order harmonics is obtained. By imposing a band-pass filter with bandwidth of 50 order harmonics on the continuum spectrum, a sub-100 attosecond isolated pulse is generated without any phase compensation. If all the phase of the continuum spectrum is properly compensated for, an isolated 9 attosecond pulse with clean profile can be obtained. Considering the practicability in experiment, we investigate the HHG spectrum of He atom in combination of chirped laser field and a low-frequency laser field ($CO_2$ laser, the intensity and wavelength is $8\times10^{13}\,\mathrm{w/cm^2}$ , 9.6 μm respectively).Still an ultra-broad continuum spectrum (covers about 600 order harmonics) is obtained and with proper phase compensation, an isolated 9 attosecond pulse can be generated, which have little differences with that discussed above.

## Acknowledgements

The work is supported by the National Basic Research Program of China (Grant No.2006CB921104, 60708008), the Project of Academic Leaders in Shanghai (Grant No.07XD14030) and the Knowledge Innovation Program of the Chinese Academy of Sciences.